\documentclass[12pt]{article}
\usepackage{graphics}
\input epsf

\newcommand{\be}{\begin{eqnarray}}
\newcommand{\ee}{\end{eqnarray}}

\begin{document}
\title{
\begin{flushright}
{\large UAHEP041}
\end{flushright}
\vskip 1cm
A supersymmetric model of gamma ray bursts}
\author{L. Clavelli\footnote{lclavell@bama.ua.edu}\quad and G. Karatheodoris
\footnote{karat002@bama.ua.edu}\\
Department of Physics and Astronomy\\
University of Alabama\\
Tuscaloosa AL 35487\\ }
\maketitle
\begin{abstract}
We propose a model for gamma ray bursts in which a star subject to a
high level of fermion degeneracy undergoes a phase transition to a
supersymmetric state.  The burst is initiated by the transition of
fermion pairs to sfermion pairs which, uninhibited by the Pauli
exclusion principle, can drop to the ground state of minimum momentum
through photon emission.  The jet structure is attributed to the
Bose statistics of sfermions whereby subsequent sfermion pairs are
preferentially emitted into the same state (sfermion amplification
by stimulated emission).  Bremsstrahlung gamma rays tend to preserve
the directional information of the sfermion momenta and are themselves
enhanced by stimulated emission.
\end{abstract}

   Within the past five years well over a hundred articles have discussed the
possibility of transitions between various local minima of the effective potential
of string theory stimulated by work by Bousso and Polchinski \cite{Bousso}, Susskind
\cite{Susskind}, and Kachru et al. \cite{Kachru03}.  
In particular the phase transition between a
vacuum similar to ours with positive vacuum energy and the vacuum of exact
supersymmetry (SUSY) with vanishing vacuum energy has been treated in string theory
\cite{Kachru03}.  In this article, we discuss possible phenomenological 
manifestations of such a transition in a dense star.
 
    We take as a starting point the three experimental-theoretical indications:
\begin{enumerate}
\item We live in a world of broken supersymmetry (SUSY) where most of the
supersymmetric particle masses are at the weak scale (several hundred GeV)
or above.  Indications for this come from successful 
SUSY grand unification predications for the
$b/\tau$ mass ratio and the $\alpha_s-sin^2(\theta)$ relationship as well
as the astrophysical indications for non-baryonic dark matter.
\item In our world there is a positive vacuum energy density (dark energy)
and a negative vacuum pressure $p_{vac}=-\rho_{vac}$
leading to an acceleration in the expansion of the universe \cite{Carroll}.
\be
      \frac{\ddot{R}}{R} = - \frac{4 \pi G}{3} (\rho_{vac} + 3 p_{vac}) > 0
\ee
\item The true ground state of the universe is a state of exact SUSY
where particles and their supersymmetric partners have the same mass.
This seems to be a persistent prediction of string theory. 
We choose to consider a transition to a flat space exact SUSY where the
vacuum energy vanishes as opposed to a 
possible transition to an anti-deSitter minimum which could also be
explored and would probably be qualitatively similar.    
  
\end{enumerate}
  
    A strict consequence of accepting these three indications and the string
theory prediction that all parameters of the theory are dynamically determined
is that the 
universe will ultimately undergo a phase transition to the true ground state 
of exact SUSY. In such a situation, only the probability per 
unit time for this transition to occur 
is, at present, unknown and subject to speculation.  Such decays of the
false vacuum were discussed in some generality by Coleman and collaborators
several decades ago \cite{Coleman}.  In a homogeneous medium, once a
critical bubble of true vacuum is nucleated, it will grow without limit.
Thus, in particular, if a bubble of critical size forms in dilute matter
it will rapidly take over the universe \cite{Coleman} with the immediate 
extermination of all life.  

Plausible suggestions have been made that the phase transition to the true
vacuum might be catalyzed in dense matter \cite{Gorsky},\cite{Voloshin}
and we argue that a bubble
of true (SUSY) vacuum, once formed, would be confined 
to the the region of high matter density.  Details are presented in another
article \cite{CP}; the argument is outlined below.
Such a situation would be in line with string theory arguments 
suggesting that
the universe might have a domain structure in which different regions in
space-time might have different physical constants, different particle
masses, and even different gauge groups.

    While superstring theory is struggling to find some experimental
confirmation beyond the (already impressive) automatic incorporation
of gravity and gauge forces, the field
of gamma ray bursts, on the other hand, is one in which rapidly
expanding observational
data is, most astronomers admit, in need of additional theoretical 
insight.
The sheer enormity of the energy release in these bursts together with
their short lifetime and pronounced jet structure make it possible
that a full explanation will not be found without some type of startling
``new physics''.
One example of such speculative ``new physics''
proposals is the quark star model of Ouyed and Sannino \cite{Sannino}.

    Although the long duration gamma ray bursts (lifetimes greater than about
two seconds) have been observationally associated with supernovae, the
existence of the required explosion has not as yet been
successfully modeled in standard astrophysical monte carlos
of supernova collapse  \cite{Janka}.  An additional energy
release mechanism such as that proposed here could, therefore, be helpful.
Conceptual gaps in conventional approaches to the theory of gamma ray
bursts based on accretion disks and hydrodynamic shock waves are, at
present, temporarily filled by the terminology of a ``central engine''
and ``firecone production''.  It is not clear that the standard model
has within it an adequate energy release mechanism nor a mechanism for
the sufficient collimation of the burst.  Models for the narrow
collimation of the bursts typically involve
the acceleration away from incipient black holes of large neutral bodies 
of matter to Lorentz parameters near $100$.  The physical basis for
the requisite strong forces is not firmly established although some
speculative ideas have been put forward.  Most workers in the field
would admit that the mechanism for launching such a jet is unclear.
Another problem in the 
conventional approaches is the lack of ``baryon loading''.  Namely, the
accelerated body of matter must be largely leptonic in order for the
energy deposition to be primarily in the gamma ray range with 
relatively little
converted to kinetic energy of heavy particles or to low energy photons.       
\cite{Zhang,Rosswog}.

In addition to not having a conceptually complete energy release mechanism
or collimation mechanism, the 
conventional astrophysical approaches do not predict the primary
quantitative characteristics of the bursts except as related to free 
parameters in the theory.
These primary quantitative observations are
\begin{enumerate}
\item burst energies 
in a narrow range near
$10^{50}$ ergs.  This assumes burst collimation, otherwise the burst
energies are much greater and widely varying \cite{Rosswog}.
\item 
typical photon energies in the $100$ KeV to $1$ MeV 
range \cite{Zhang}.
\item
Burst durations of from some $20$ milliseconds to $200$ seconds with 
the duration distribution having a pronounced dip at about $2$ s 
\cite{Kouveliotou}.
\end{enumerate}
This, however, is not to say that no progress is being made in 
exploring conventional astrophysical possibilities.  For instance,
several models
have been proposed in \cite{Lee}.  In one of these it is suggested
that about $10^{50}$ ergs (depending on accretion disk viscosities
and an assumed efficiency of $1 \%$)
could be converted from $\nu \overline{\nu}$ into an $e^+e^-$
plasma which could then be made available for the production of
a relativistic fireball.  This model could be in line with the
``cannonball'' model of Dar and DeRujula \cite{DeRujula} which
involves the acceleration of a relativistic fireball away from a 
progenitor star but the nature of the central engine and jet
launching mechanism are still uncertain.
 
    Given the magnitude of the long-standing challenge posed by 
gamma ray bursts, we would hope that,
while the feasibility of mechanisms such as the above 
is being investigated, broad latitude is also given to 
the discussion of ideas beyond the standard model
even if they are necessarily less fully developed 
and seemingly more speculative.

     In the current paper we propose as a model for the ``central 
engine'' the lifting of Pauli blocking due to a SUSY phase 
transition.  The resulting energy release could be
utilized in a subsequently conventional astrophysical
model for the gamma ray bursts.  However, we note that the 
transition to a largely bosonic final state also suggests a 
natural mechanism for the burst collimation.      
 
     Our proposal is based on the following scenario.

\begin{enumerate}
\item  In a region of space with a high level of fermion
degeneracy there is a phase transition to a supersymmetric
ground state.  In the SUSY phase, electrons and their SUSY
partners (selectrons) are degenerate in mass as are the
nucleons and snucleons, photons and photinos etc.  
A critical assumption for the current work is that the common
mass of electrons and selectrons in the exact SUSY phase is
no greater than the electron mass in our broken SUSY universe.
This assumption is supported by string theory which 
predicts massless ground
state supermultiplets in the true-vacuum, exact SUSY phase.
In addition, one can note that, in the popular model of
radiative breaking of the electroweak (EW) symmetry, SUSY
breaking and EW breaking are linked so that in the absence
of susy breaking, the ground state supermultiplets are massless.
We know of no calculation in the literature requiring a
necessarily higher common ground state mass.
We assume for definiteness and simplicity an equality of 
the common mass in the exact SUSY phase and the particle
mass in the broken phase.
   
\item In the SUSY phase, electron pairs undergo
quasi-elastic scattering to selectron pairs which, uninhibited
by the Pauli principle, can fall into the lowest energy state
via photon emission.  These photons are radiated into the
outside (non-SUSY) world.  Other photons are emitted at the
boundary to conserve momentum as the selectrons are reflected
by the domain wall not having sufficient energy
to cross into the non-SUSY domain.  A highly collimated
jet structure could be produced by the stimulated emission of
sfermions and photons. 
\item
Simultaneous with electron conversion 
into selectrons, nucleons within heavy nuclei convert into snucleons.
With no further support from the electron
degeneracy, the star collapses to nuclear density under
gravitational pressure.
\item  Remaining nucleon pairs then undergo the analogous
conversion to snucleon pairs with the cross section mediated
by the strong exchange of supersymmetric pions.  This process can be 
temporarily interrupted by brief periods of fusion 
energy release but then continues until the star falls below 
the Schwarzschild radius and becomes a black hole, thus
extinguishing the gamma ray burst if it has not already ended.
The exact behavior of a SUSY bubble in a dense star is, obviously,
a complicated
problem and only the simplest zeroth order calculations are within
the scope of this initial paper.
\end{enumerate}

    In this model bursts could be due
to the decay of isolated white dwarfs which are 
absolutely stable in standard astrophysics.  We therefore predict 
the existence of low mass black holes below the Chandrasekhar 
limit.  In the following we show, in outline form,
that the mechanism produced here
can quantitatively, though roughly, account for the observations of 
stellar explosion, total energy release, minimum
burst duration, average photon energy, and jet collimation.
No other comparably parameter-free model predicting these 
primary quantitative features of the bursts exists at present.
A more rigorous modeling of the burst in the SUSY phase
transition framework, addressing some of the
secondary characteristics, is deferred to a later
paper and to future investigations.

    Transitions between vacua of differing amounts of supersymmetry 
have been considered in string theory \cite{Kachru} and lie at the
basis of string landscape models.  In order for such phase transitions
to occur, the effective potential must be dynamically determined as
in string theory and some other models of spontaneous SUSY breaking.  In a
model such as the Minimal Supersymmetric Standard Model (MSSM) where
the SUSY breaking is attributed to fixed parameters, one would not
expect phase transitions between vacua with differing amounts of 
supersymmetry. 
Catalysis of vacuum decay by matter effects has been rigorously
treated in two dimensions \cite{Gorsky}. This catalysis 
is more difficult to treat in four dimensions but we adopt the
idea that the SUSY transition will be much more
likely to nucleate in a dense star than elsewhere in space. 
One likely manifestation of this catalysis might be that the critical
radius above which a SUSY bubble will expand and below which 
be quenched is much greater in vacuum than in dense matter.
From the expression of ref.\cite{Coleman} for the vacuum case, we
would expect a critical radius of
\be
      R_c = \frac{3 S}{\epsilon + \Delta \rho}
\ee
where $\epsilon + \Delta \rho$ is the ground state energy density in 
the broken
SUSY phase minus the ground state energy density in the exact SUSY
phase and $S$ is the surface tension of the bubble.  
Here, $\epsilon$ is the observed vacuum energy density and
$\rho$ is the ground state matter density.  The difference $\Delta
\rho$ is the excitation energy density in the broken SUSY phase.
For the nominal white dwarf ignoring density inhomogeneity, the
kinetic energy density of the degenerate electron gas is about
\be
    \Delta \rho \approx 6 \cdot 10^{34} {\displaystyle MeV/m}^3 .
\label{deltarho}
\ee
Inhomogeneity effects are the subject of an article currently
in preparation.
It has been argued \cite{Frampton} that the current longevity of the
universe requires that $R_c$ in vacuum be greater than 
the galactic radius.  Although he did not consider supersymmetry
specifically, his analysis suggests a lower limit on $S$.
\be
      S > \frac{R_{galaxy} \epsilon}{3} = 5.6 \cdot 10^{23}
    {\displaystyle MeV/m}^3
\label{Sestimate}
\ee
Extrapolating to a dense medium from the vacuum calculation of
ref.\cite{Coleman}, the transition probability per unit time in 
a homogeneous body of volume $V$ is expected to be of the form 
\be
      \frac{1}{N} \frac{dN}{dt} = A V e^{-(\frac{\tilde \rho}{\epsilon + 
\Delta \rho})^3}
\label{transprob}
\ee
where, in the vacuum, $\Delta \rho = 0$  \cite{Coleman}.  In the 
simplest cases,
$\tilde \rho$ is proportional to the $4/3$ power of the surface tension
which is usually treated as a constant but could be density dependent
at high density.
The exponential factor grows rapidly with $\Delta \rho$ up to $\tilde\rho$ and
then saturates.  For more dense systems the transition rate is 
proportional to the volume.  The parameter $A$ is at present undetermined.
If $\tilde\rho$ is of order the nominal white dwarf electron kinetic energy 
density of eq.\ref{deltarho}, the other parameters
can be reasonably chosen so that the transition probability in vacuum and
the transition probability in a heavy nucleus are negligible while the
rate in a dense star is appreciable.  In this case we would predict bursts
from isolated white dwarf stars and from more massive collapsing objects
as they approach white dwarf density.  Depending on the value of $\tilde\rho$,
there could also be significant transitions in Neutron stars.  
Clearly, at present the
rate of SUSY transtions is somewhat parameter dependent but, as we will
show, the zeroth order manifestions of such a transition in a white dwarf star, 
once it occurs, are relatively unique.  

     If a SUSY bubble forms in an electron gas, electron pairs will
convert to selectron pairs.
\be
       e^{-}e^{-} \rightarrow {\tilde e}^{-}{\tilde e}^{-}
\label{sigma}
\ee 
The cross section for process \ref{sigma} is, apart from logarithmic factors, \cite{Keung}
\be
     \sigma_0 = \frac{\pi \alpha^2 (\hbar c)^2}{4 <E>^2}  .
\ee
Thus, the half life of a sample of electrons undergoing this process followed 
by bremstrahung is
\be
     \tau \approx \frac{1}{\alpha \sigma_0 \rho v} \approx 
           \frac{16 \pi <E>^3 \hbar}{\alpha^3 (c p_{max})^4}
         \approx 3.3 \cdot 10^{-13} s  
\ee
where we have borrowed parameter values from considerations below.
Once the radiated photons have left the bubble,
the broken-SUSY phase can no longer quench the SUSY bubble
since the sparticles are prohibitively massive in the 
normal world.

    For the bremstrahlung to occur before the bubble collapses
thus trapping the selectrons, the minimum size of the bubble must,
therefore, be roughly of order
\be
     r > \frac{c}{\alpha \sigma_0 \rho v} \approx 10^{-4} m  .
\ee
The resulting constraint on the surface tension is well within 
that suggested by eq. \ref{Sestimate}. 

    If we consider the transition as beginning with the
strong transition from nucleons to snucleons, this minimum
bubble size might be a few orders of magnitude smaller but still 
much greater than nuclear size.
The volume factor in eq. \ref{transprob} makes it highly 
unlikely that the SUSY transition will
take place in a terrestrial heavy nucleus but we postulate
that the process occurs
in fermi degenerate stars with a probability per
unit time fixed by the rate of gamma ray bursts divided by
the number of such stars.  In the current state of the art
with respect to vacuum decay we cannot calculate this
probability nor do we need to know its value for our 
present considerations.  Nevertheless, we can note that
estimates of
the number of white dwarfs in our (typical) galaxy are of
order $10^{9}$.  The number of gamma ray bursts per
year per galaxy is about $5 \cdot 10^{-7}$ assuming a
$5^\circ$ burst opening angle.  Thus, if the
SUSY phase transition model for the bursts is correct, 
the probability for a given white dwarf star to explode
in a given year is less than $10^{-15}$. 
Until the
phase transition takes place, the white dwarf will cool
according to standard physics.  Thus, even if the
estimate of white dwarf numbers or burst rates are
off by some orders of magnitude, the
present model is clearly not in conflict with 
current observations of white dwarf cooling.  
It is also possible that many, or even most, of the
SUSY phase transitions result only in a neutrino burst
with the gamma rays being swallowed by the subsequent
black hole.  Even then, it is still highly improbable that 
a  particular white dwarf would be observed to
suddenly disappear.  In this connection one could note that
there is, in fact, a long-standing shortage of cool white
dwarfs \cite{Winget} and, perhaps, a surplus of dark objects
of white dwarf mass \cite{Oppenheimer}.   The MACHO experiment
has also detected a surprisingly large number of dark objects
of low mass \cite{MACHO} which, in the SUSY model,
could be interpreted as SUSY black holes of mass below the
Chandrasekhar limit.  A repeat of these observations with
increased sensitivity is highly desirable.

     We consider the case of a typical white dwarf of solar mass
($M=1.2 \cdot 10^{60}$Mev/c$^2$) and earth radius ($R = 6.4 \cdot 10^6 m$)
supported as in the standard astrophysical model by electron
degeneracy.  That is the number of electrons with momentum between
$p$ and $p+dp$ is
\be
         dN = \frac{8 \pi p^2 dp V}{(2 \pi \hbar)^3}
\ee
with
\be
         p_{max}= \left ( \frac{3N}{8 \pi V} \right )^{1/3} 2 \pi \hbar
                    = 0.498 \displaystyle{MeV/c}  .
\label{pmax}
\ee
Here, N is the total number of electrons in the white dwarf
\be
      N = 6 \cdot 10^{56}
\ee
where we have assumed equal
numbers of electrons, protons and neutrons.
The average squared three-momentum of the electrons is
\be
      <p^2> = \frac{3}{5} {p_{max}}^2
\ee
and the average electron energy is
\be
     <E> = mc^2 \left ( \sqrt{(1 + <p^2>/(mc)^2)} + g \right )  .
\ee
g is a small correction term given by
\be
    g = \sum_{l=0}^{\infty} (\frac{p_{max}}{mc})^{2l+4}\frac{\Gamma(3/2)}{(l+2)!
          \Gamma(-1/2-l)}\left( \frac{3}{7+2l} - (3/5)^{l+2} \right)  .
\ee
When the final state selectron comes to rest after bremstrahlung or reflection at the
boundary the energy release per electron is
\be
     \Delta E = <E> - mc^2 = .11 {\displaystyle MeV}  .
\ee
This photon energy is in the gamma ray range as observed in the bursts.
The total energy release from all electrons is
\be
     N \cdot \Delta E = 1.2 \cdot 10^{50} {\displaystyle ergs}  .
\ee
The half life of a sample of electrons undergoing process \ref{sigma} is
\be
     \tau \approx \frac{1}{\sigma_0 \rho v} \approx 2.4 \cdot 10^{-15} s .
\ee
Since this is essentially instantaneous, the time scale of the selectron burst is
fixed by the time it takes for the SUSY phase to spread across the star and for the
photons from the far side of the star to traverse the star.  The speed of light 
gives a lower limit to the duration of a burst from the nominal white dwarf.
\be
      \tau \approx \frac{R}{c} \approx 0.02 s  .
\ee
This is roughly the observed minimum duration of the gamma ray bursts.
However, this prediction is complicated by the fact that the bubble expansion
speed in dense matter might be significantly slower than the speed of light.
Using the average density, the speed of sound in the nominal white dwarf would
lead to a bubble growth time of $2$ s. 
We have, however, not taken into account
the variations in radii among white dwarfs.  In addition, one needs to
consider the varying free collapse time discussed below of a star
relieved of Pauli blocking. The investigation of these and 
many other possible effects relevant to the duration 
distribution of the bursts in the phase transition model is at an early stage.  
In the standard astrophysical approaches to gamma ray bursts, the
duration distribution is also in early stages of understanding.  
Similarly, the rapid time variability or ``spikey'' nature of the bursts
presents challenges to both the phase transition and conventional approaches.
In the phase transition model these spikes could be due to emission from different
momentum levels in the degenerate electron sea or to other quantum decoherence effects.
In the conventional approach, the spikes are often attributed to ``sub-jets'' within
the burst although their physical origin cannot be determined without a full theory
of the central engine.  

During the conversion of electrons, the lifting of electron degeneracy causes the
star to collapse rapidly under the gravitational forces until nuclear density is
reached.  Until then, however, separated nuclei are outside the range of strong
interactions so nucleon conversion proceeds only within individual nuclei.
Initially SUSY conversion within nuclei occurs via the strong reactions
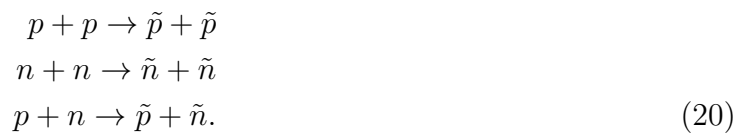
\be\nonumber
     p + p \rightarrow {\tilde p} + {\tilde p}\\ \nonumber
     n + n \rightarrow {\tilde n} + {\tilde n}\\ 
     p + n \rightarrow {\tilde p} + {\tilde n}  .
\label{nucleonconversion}
\ee
These processes are mediated by pioninos (the SUSY partners of the pions).  
In a white dwarf the dominant nuclei are Carbon and Oxygen.  We can estimate
the energy release in the processes \ref{nucleonconversion} using a simple
three dimensional square well model.  After SUSY conversion to bosonic
particles, the shell model excitation energy will be released.  
Using a Carbon radius of $2.3$ fm \cite{Krane}
, we estimate that there will be $3.0$ MeV released per Carbon nucleus
for a total energy release in the nominal white dwarf of $4.9 \cdot 10^{50}$
ergs.  This is slightly greater than that
found from the electron sea.  
 
    If there are appreciable amounts of odd isotopes,
the SUSY transition will not go completely within separated
nuclei and, relieved of the electron degeneracy, the star will collapse 
under gravitational pressure until the remaining
protons and neutrons achieve fermion degeneracy at, we assume, nuclear 
densities ($(\frac{N}{V})^{1/3}=0.47$ fm$^{-1}$ \cite{Preston}).
At nuclear densities, the remaining nucleons will undergo SUSY conversion to scalar particles
with further release of energy after which time the star will collapse to a black hole.
Thus the SUSY phase transition model is a multi-component model.
Because of the high mass of nucleons and their non-relativistic velocities, the nuclear
energy release may not contribute significantly to the collimated burst but may 
contribute to the afterglows.   

    Classically, if a piece of a star of mass $\Delta m$ implodes from a radius 
$r_0$ to a radius $r$, its final kinetic energy will be
\be
    \frac{1}{2} {\Delta m} \left( \frac{dr}{dt} \right)^2=
      {\Delta m} \left(\frac{GM}{r} - \frac{GM}{r_0} \right) .
\ee
A freely imploding star of initial radius $r_0$ at time $t=0$ will
have at time $t$ a radius $r$ given by
\be
     \theta + \frac{\sin(2 \theta)}{2} = t \quad \sqrt{\frac{2GM}{r_0^3}}
\ee
where
\be
    \theta = \tan^{-1}(\frac{r_0}{r}-1)^{1/2}  .
\ee
If, as will always be the case, the initial radius is far greater
than the final radius, the collapse time, assuming complete lifting
of the Pauli blocking, will be
\be
     t = \frac{\pi}{2} \left( \frac{8 \pi G \rho}{3} \right)^{-1/2}
\ee
where $\rho$ is the initial density.
This can be written
\be
     t = 1.53 s \left( \frac{\rho}{\rho_{WD}}\right)^{-1/2}
\label{collapsetime}
\ee
where $\rho_{WD}$ is the typical white dwarf density 
(solar mass, earth radius).  

    Although further study is needed, it is tempting to suspect 
that this time is related
to the observed dip at $2$ s in the burst duration distribution.
Objects with a natural burst duration near $2$ s might have only
a partial SUSY conversion before gravitational collapse thus
resulting in a build-up of events at lower burst times.  
As can be seen from eq.\ref{collapsetime}, a transition in a
star of lower density will have a longer collapse time.
In addition, as the star approaches the Schwarzschild radius,
general relativistic effects are
expected to stretch out the collapse time and red-shift the
final stages of afterglow.  Other sources of afterglow are
irradiated circumstellar material.

    In conventional astrophysical models for the bursts, the 
duration distribution is often assumed to come from a viewing 
angle dependence although the existence and location of the dip 
is not easily predicted \cite{Zhang,Yamazaki}.
     
    Next we explore
the suggestion that the strongly collimated jet structure is due to a bose
enhancement of the emitted selectrons, sprotons, and
bremstrahlung photons, i.e. a stimulated emission.  The matrix
element for the emission of a selectron pair with momenta $\vec{p}_3$ and
$\vec{p}_4$ in process
\ref{sigma} in the presence of a bath of previously emitted pairs is
proportional to
\be
\nonumber
     {\cal M} \sim <n(\vec{p}_3)+1, n(\vec{p}_4)+1\mid a^\dagger(\vec{p}_3)
     a^\dagger(\vec{p}_4)\mid n(\vec{p}_3), n(\vec{p}_4)> \\
 \sim \sqrt{(n(\vec{p}_3)+1)}
     \sqrt{(n(\vec{p}_4)+1)}  .
\ee
The cross section is, therefore, proportional to
     $(n(\vec{p}_3)+1)(n(\vec{p}_4)+1)$ .
The full modeling of this enhancement requires a multi-
dimensional monte carlo (three integrations for each initial state
electron plus two angular integrals for one of the final state
selectrons although two of these integrals can be done trivially).  
We would also need the cross section for process
\ref{sigma} without neglecting the electron mass.
This complete calculation has been 
recently published \cite{CP}.
Here we content ourselves with the following
statistical model which has no dynamical input but
provides a simplified
demonstration of the principle of stimulated emission of Bosons.

     We generate events in the three dimensional space of one of
the selectrons momentum magnitude, $p_3$, polar angle cosine,
$\cos(\theta_3)$, and azimuthal angle, $\phi_3$, assuming
that each event takes place in the $CM$ system.  Then
$\vec{p}_4=-\vec{p}_3$ and $n(\vec{p}_4)=n(\vec{p}_3)$.
Initially all the $n's$ are zero but once the first transition has
been made populating a chosen $\vec{p}_3$, the next transition is
four times as likely to be into the same state as into any other
state.  Because of the huge number of available states, the second
transition is still not likely to be into the same $\vec{p}_3$
state, but as soon as some moderate number of selectrons have been
created with a common $\vec{p}_3$, the number in that state
escalates rapidly, producing a narrow jet of selectrons.  These
selectrons decay down to the ground state via bremstrahlung
photons which are also Bose enhanced leading to a narrow jet of
photons which can penetrate the transparent domain wall and
proceed into the non-SUSY phase.

      We model this simplified process by standard monte carlo
techniques.  
To deal with the three dimensional space we define a composite
integer variable, k, defined as
\be
     k = n_{bin}^2 n_1 + n_{bin} n_2 + n_3
\ee
where $n_{bin}$ is the number of bins in each of the three variables, 
$p_3$, $\cos(\theta_3)$, and $\phi_3$.  The $n_i$
are integers running from 0 to $n_{bin}-1$ and are related to the three variables by
\be
\nonumber
     p_3 &=& p_{3,max} (n_1 + 1/2)/n_{bin}\\
\nonumber
     \cos(\theta_3) &=& (2 n_2 + 1)/n_{bin} -1 \\
     \phi_3 &=& \pi (n_3 + 1/2)/n_{bin}  .
\ee
k runs from 0 to $n_k = n_{bin}^3 - 1$ and each value of k corresponds
to a unique value of the three variables, $p_3$, $\cos(\theta_3)$, and $\phi$.
At each stage in which there are some
occupation numbers $n(j)$ we calculate the normalized sum
\be
     R(k) = \frac{\sum_{j = 0}^{k}(n(j)+1)^2}{\sum
      _{j=0}^{n_{k}}(n(j)+1)^2}  .
\ee
$R(k)$ is a monotonically increasing function of bin number $k$, varying
between $0$ and $1$. 
    
      Then choosing a random number $r$ between $0$
and $1$, if $r<R(0)$ we add an event to the first bin
and repeat the process.
If $r > R(k)$ and $r \leq R(k+1)$ we add an event to
bin $k+1$ and repeat the process.
After $10^5$ events (still a
tiny fraction of the available $10^{56}$) we arrive at the
distribution shown in table 1 with $n_{bin}=10$ and
$p_{max}=0.498$ MeV/c as in eq.\ref{pmax}.

    This toy model gives, of course, no insight into the actual
width of the jets since no dynamics is incorporated.
In addition, the photon energy is here taken to be the
full kinetic energy of the produced sparticle neglecting multiple
bremstrahlung effects etc.  A more physical picture of the
jet distributions should come out of the more complete dynamical
monte carlo to be treated in the near future.  

\begin{table}[tbp]
\begin{center}
\begin{tabular}{||cc||cc||cc||}
\hline
         &    &                &   &        &   \\
 $p_3$(MeV) & N  & $\cos(\theta_3)$ & N & $\phi_3$ &  N\\
         &         &           &         &         &    \\
   0.02  &     52  &   -0.900  &     50  &  0.157  &    33\\
   0.07  &  99608  &   -0.700  &     60  &  0.471  &    23\\
   0.12  &     32  &   -0.500  &     34  &  0.785  &    49\\
   0.17  &     35  &   -0.300  &     71  &  1.100  &    49\\
   0.22  &     58  &   -0.100  &     45  &  1.414  &    44\\
   0.27  &     52  &    0.100  &  99598  &  1.728  &    48\\
   0.32  &     31  &    0.300  &     22  &  2.042  &    46\\
   0.37  &     49  &    0.500  &     33  &  2.356  & 99604\\
   0.42  &     30  &    0.700  &     49  &  2.670  &    65\\
   0.47  &     54  &    0.900  &     39  &  2.985  &    40\\
         &         &           &         &         &    \\
\hline
\end{tabular}
\caption{Development of jet structure in a simplified statistical 
model.  The first column gives the photon energy,
the third gives the polar angle cosine, and the fifth gives the 
azimuthal angle.  The second, fourth, and sixth columns give
the number of photons in the first 100,000 with those values of
$p$, $\cos(\theta)$, and $\phi$.}
\end{center}
\end{table}

    We have presented a physical picture that, accepting
its premise, does lead to an explosion into a burst of gamma rays 
of near MeV energies, with a pulse duration ranging down to a 
small fraction of a second,
highly collimated in angle, and containing a total burst energy
of about $10^{50}$ ergs.  The SUSY phase transition takes place
preferentially at high density.  It is not clear whether 
isolated stars have sufficiently high density over sufficiently
large volumes or whether accretion plays an important role
in providing these necessary conditions.  In the latter case
the SUSY star model could be incorporated into
the standard astrophysical approaches as a model for the 
central engine.

      Although many details of the SUSY phase transition model
remain to be explored, the gross features of the observed bursts are 
relatively easily understood with one radical, though not unwarranted,
assumption but no free parameters.  Given the existing physical basis 
for our assumption we do not regard the present hypothesis as overly
speculative.  The model leaves open the
question whether evidence for similar SUSY phase transitions can be
observed elsewhere in astrophysics or in terrestrial experiments such 
in heavy ion collisions.

{\bf Acknowledgements}

    This work was supported in part by the US Department of Energy under 
grant DE-FG02-96ER-40967.
We gratefully acknowledge discussions with Cecilia
Lunardini of the IAS, Princeton and with Jerry Busenitz, Phil Hardee, 
Ray White, Enrique Gomez, and Irina Perevalova  at the University of Alabama.


\end{document}